# Knowledge Co-creation in the OrganiCity: Data Annotation with JAMAiCA

Aikaterini Deligiannidou, Dimitrios Amaxilatis,
Georgios Mylonas
Computer Technology Institute & Press "Diophantus"
Rio, Patras, Greece, 26504
{kdeligian, amaxilat, mylonasg}@cti.gr

Evangelos Theodoridis
ICRI Sustainable Connected Cities
Intel Labs Europe
London, UK
evangelos.theodoridis@intel.com

*Abstract*— Numerous smart city testbeds and system deployments have surfaced around the world, aiming to provide services over unified large heterogeneous IoT infrastructures. Although we have achieved new scales in smart city installations and systems, so far the focus has been to provide diverse sources of data to smart city services consumers, while neglecting to provide ways to simplify making good use of them. We believe that knowledge creation in smart cities through data annotation, supported in both an automated and a crowdsourced manner, is an aspect that will bring additional value to smart cities. We present here our approach, aiming to utilize an existing smart city deployment and the OrganiCity software ecosystem. We discuss key challenges along with characteristic use cases, and report on our design and implementation, along with preliminary results.

## Introduction

Smart cities have slowly been turning from a vision of the future to a thing of the present, through the efforts of numerous research projects, technology startups and enterprises, combined with the recent advancements in informatics and communications. It is currently a very active field research-wise, with a lot of work dedicated to developing prototype applications and integrating existing systems, in order to make this move from a vision to reality.

Although a lot of emerging technologies in the smart city context still compete in the same space, a number of actual use-cases and methodologies have surfaced in multiple smart city instances. For example, much buzz has been generated around the smart city IoT testbed and experimentation concept, like in the case of SmartSantander [11]. Another example is the utilization of open data portals in smart cities, like CKAN [7] an open source solution provided by a worldwide community and Socrata [8] an enterprise solution backed up by a an IT company. Additionally, technologies like MQTT and CoAP are frequently utilized in recent smart city research projects to provide real-time communication with the deployed infrastructure, and progress towards becoming Internet standards.

However, there remain essential answers to be found revolving around a central question: what do we do with all of these data collected, and how do we make sense out of them by extracting knowledge, i.e., something actually useful, going beyond a technology demonstrator? Moreover, how do we provide usefulness to citizens and how do we involve them in the smart city lifecycle or engage them in the whole process?

Essentially, it gets to the point of asking how do we actually make a city smarter, and the definition of a smart city itself. We believe part of the answer to this question lies in creating more "useful" information out of raw sensors or other kind of data representing observations of the urban environment. For example, certain events generate data reported by the city sensing infrastructure, but are, more often than not, missing an appropriate description. Consider the case of a traffic jam inside the city center; it generates sensed values in terms of vehicles' speed, noise and gas concentration. Moreover, in most cases, such sensed values are reported by multiple devices or services while missing useful correlations in the data streams. We believe that adding data annotations to smart city data through machine learning mechanisms or crowdsourcing mechanisms, can help reveal a huge hidden potential in smart cities.

In this work, we discuss the design and implementation of *JAMAiCA* (Jubatus Api MAChine Annotation), a system for aiding smart city data annotation through classification and anomaly detection, which is currently being employed in the *OrganiCity* project ecosystem. On the one hand, it aims to simplify the creation of more automated forms of knowledge from data streams, while on the other hand it serves as a substrate for crowdsourcing data annotations via a large community of contributors that participate in the knowledge creation process. We strongly believe that communities like data scientists, decision makers and citizens should get involved in deployments of Future Internet systems, for them to be practical and useful.

Regarding the structure of this work, we first report on previous related work, and continue with a discussion on challenges associated with knowledge creation in smart cities. We then present a small set of use-cases to highlight how our system relates to this vision. We continue with a presentation of our design and system architecture, complemented with a description of our current implementation and some preliminary results we have produced so far.

We now proceed with a short introduction to the OrganiCity project and the ways our system relates to its overall ecosystem.

### A. OrganiCity and Co-creation

OrganiCity aims to engage people in the development of future smart cities, bringing together three European cities: Aarhus (Denmark), London (UK) and Santander (Spain). Co-creation with citizens is its fundamental principle, i.e., defining novel scenarios for more people-centric applications inside smart cities, exploiting the IoT technologies, heterogeneous data sources and using enablers for urban service creation and IoT technologies. It places *organicitizens* (i.e., citizens actively involved in the project) at the heart of its life cycle, both by helping shape application use-cases and by simplifying ways to contribute data or other resources.

Fig. 1 provides an overview of OrganiCity. The project aims to provide an Experimentation as a Service (EaaS) platform, i.e., it is designed to provide data streams from diverse sources inside a smart city to various "consumers", like IoT experimenters, SMEs, municipalities, etc. At the same time, it aims to allow the participatory engagement of communities in co-creating urban knowledge. This will be done by means of end-user applications that provide meaningful representations of the produced smart city data, and also "tools" that will allow these end-users to make their own contributions.

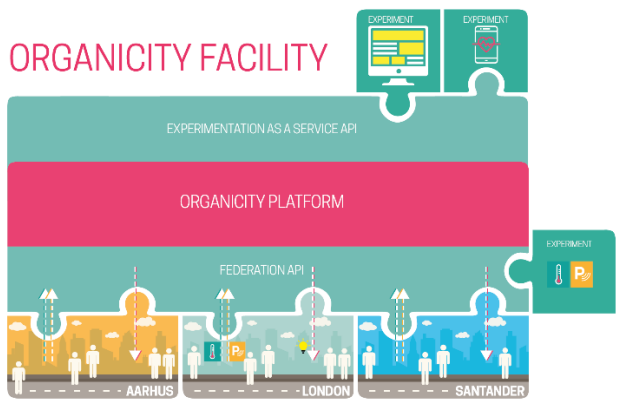

Figure 1: Overview of OrganiCity

*JAMAiCA* is meant to act as both an end-user tool and a service for other applications to further extend the data annotation functionality of OrganiCity. We follow an open-source approach for the implementation of our system as we believe that this will further entice more tech-savvy citizens to engage actively in future revisions of the system.

## PREVIOUS WORK

Although there have been pioneering studies and applications on combining human and machine intelligence, it is believed that research in this field is still at its infancy stage. [10] presents a vision on the potential of combination patterns of human and machine intelligence, identifying three possible patterns *sequential*, *parallel* and *interactive*. Moreover, in [12] authors present a crowd-programming platform that integrates machine and human based computations. Their system integrates mechanisms for challenging tasks like human task scheduling, quality control, latency due to human behavior etc.

SONYC [19] is an example of a project with a very specific use-case, employing machine learning algorithms to classify acoustic readings into various types of noise encountered inside an urban environment. It is a very interesting approach, with similarities to our vision of providing a generic substrate to simplify the process of knowledge extraction and data annotation contributions. Moreover, learning from the crowds, by using the crowdsourced labels in supervised learning tasks in a reliable and meaningful way is invegistated in [17, 18].

Taxonomies are ubiquitous in organizing information, by grouping digital objects/content to categories and/or mapping them to abstract concepts expressing meanings, entities, events etc. Most of the modern social networking applications (like Flickr) or online collaborative tools (like Stack-Exchange) are relying heavily on an underlying taxonomy. Building and curating a taxonomy is a challenging task that requires deep knowledge of the data characteristics. Taxonomies are usually created by small groups of experts and target a very specific application domain. Folksonomies are quite popular in online applications and they are categories of tags collectively organized by the users of the applications. Such taxonomies usually have weaknesses like double entries, misclassified tags, entries with typos or ambiguities in the categories, and so on. By processing in the background, it is possible to normalize folksonomies by mapping categories to knowledge bases (like Wikipedia or WordNet).

In [13], the authors propose a workflow that creates a taxonomy from collective efforts of crowd workers. In particular, the taxonomization task breaks down into manageable units of work and an algorithm coordinates the work. The algorithm takes every item and solicits multiple suggested categories for it, from different workers. A new set of workers then votes on the best suggested category for each item. Afterwards, workers need to consider every item with all of these 'best' categories and judge their relevance. These data are later used to eliminate duplicate and empty categories, and to nest related categories. Results show that the produced taxonomy is competitive in quality and price compared to expert information architects, although the adoption of machine learning approaches could optimize the categorization process. Although that taxonomies and tagging of objects with keywords of the taxonomy, the problem is, that there is no common agreement about the semantics of a tagging, and thus every system uses a different representation. An effort for the development of a common tagging ontology with Semantic Web technologies is described in [14].

Designing and developing smart cities is a concept that has drawn tremendous attention from the public and the private sector. Each one of the scientific disciplines like urban engineering, computer science, sociology and economics, provide unique perspectives on making cities more efficient. In most of these cases, multidisciplinary approaches are required to tackle complex problems. A large number of projects are trying to leverage modern information and communication technologies, like IoT/Future Internet and the semantic web, in order to build novel smart city services and applications. An example is the SmartSantander project [11], that has developed one of the largest Future Internet infrastructures globally, located at the center of the city of Santander in Spain. A well-established city-wide IoT experimentation platform that moved testbeds from labs to the real world and that offers experimentation functionality, both with static and mobile deployed IoT devices, together with smartphones of volunteers inside the urban areas. Another example is CitySDK [14] that tries to harmonize APIs across cities and provide guidelines about how information should be modeled, propose ways that data should be exchanged and how services and applications should be designed and developed. The project benefits from semantic web technologies and focuses on application domains like citizen participation, mobility and tourism.

Given that cities are dynamic and evolving ecosystems, there is a need to continuously link, interpret and utilize information before it is outdated. Therefore, real-time data processing and stream-based annotation is a critical endeavor that has to be dealt by smart city frameworks. CityPulse [3] introduces a framework for real time semantic annotation of streaming IoT and social media data to support dynamic integration into the Web. The framework employs a knowledge-based approach for the representation of the data streams and uses the Advanced Message Queuing Protocol (AMPQ) to increase the communication performance of the system, as the amount of data generated by IoT devices can be enormous. It also presents a lightweight semantic model to represent IoT data streams, built on top of well-known models, such as TimeLine Ontology, PROV-O, SSN and Event Ontology. In terms of creating high-level concepts from the large amount of data produced, another similar approach has been carried out in [9]. In both semantic representation frameworks, pattern construction is performed using the Symbolic Aggregate Approximation (SAX) technique. The approach in [9], introduces a method to automatically create a semantic ontology, without requiring preliminary training data, using an extended k-means clustering method and applying a statistical model to extract and link relevant concepts from the raw sensor data. The framework can be used in control and monitoring applications that use the sensory data to observe the status of a physical entity or to provide an overall view of the changes and related occurrences over a period of time, but it does not for real-time processing. Finally, in [16] the authors propose principles for semantic modelling of city data.

DATA ANNOTATION IN SMART CITIES – CHALLENGES

In this section, we briefly discuss a set of key challenges regarding data annotation in smart cities. All of them need to be addressed in the near future, in order to enable a more engaging and secure experience for citizens/contributors on the one hand, and to produce a more meaningful result from the systems' side on the other. We focus on this set of specific challenges, due to their importance and our experience from OrganiCity.

*Privacy and overall security issues* are a central challenge in the context discussed here. Consider the case of a volunteer taking noise level measurements along his daily commute, or being tasked to add annotation contributions by a smart city system based on proximity to certain events. Even in such simple scenarios, anonymization techniques should be used to ensure that neither personal data, nor interactions are revealed.

Another important issue is the *correlation of different types of smart city data* that can potentially point to the same event, in other words, how to facilitate knowledge extraction through such data. We currently have data produced by IoT infrastructure installed inside city centers. However, there is relatively small research focus on discovering relations between these data, e.g., if noise level readings are related to data referring to a live concert event, or can be attributed to a single event as *results*, being produced by a specific situation taking place somewhere inside the city.

Moreover, there is the issue regarding the *nature of data available in smart city data repositories*, being data inserted by humans or IoT infrastructures. Both sources can be unreliable, or even malicious. With respect to sensing infrastructure, we also have the issue of the hardware malfunctions, as well as spatiotemporal effects on the data produced. In most cases, the hardware utilized aims for large-scale deployments, thus being not so accurate or having calibration issues. Additionally, environmental conditions, e.g., excessive temperature or humidity, may have an effect on the sensitivity of the sensing parts. The issue is how to produce data annotation based on such an infrastructure, which can function with a varying degree of credibility during a single day. Reputation mechanisms are an example of measures that can aid in this direction, either human or machine-based, in order to filter out less reliable data sources.

The issue of *end-user engagement* with respect to data annotation and knowledge extraction is, in our opinion, another major challenge. We also think that user contribution is twofold: end-users can contribute to a smart city system by adding data annotations, but also contribute data through incentivization or gamification. Although most current crowdsourcing platforms utilize a desktop or web interface, the crowdsourcing of data annotations does not have to be limited to that. It can also be performed through smartphones and be incorporated to the user's everyday life. The interaction of end-users through such a tool could help relate in a more personal way and help maintain the interest in participating. Moreover, annotation of events or sensed results could be more interactive and focus at users, or even user groups, near the actual space of the event in question.

Smart city facilities usually integrate a *large number of data sources of various types* sharing observations for environment, air quality, traffic, transport, social events and so on. These data sources might be static (they are not streaming data and have a fixed value until they are updated by an offline process) or might be dynamic (streaming data constantly). Building a taxonomy on this multi thematic environment is not straight forward as some subcategories of tags might be shared between different types of data sources and other might be orthogonal. Moreover, as the dynamic data sources have a temporal dimension, annotations might characterize the overall behavior and observations of the data sources or observations falling into a specific time interval. Furthermore, as data sources might be mobile (e.g. an IoT device on a bus or a smartphone) an annotation might characterize a specific location inside the city and for a specific time interval. Embedding in the taxonomy these spatiotemporal characteristics introduces new requirements and extensions to the traditional methods. Standards like W3 Web annotation data model and protocols[1] do not cover sufficiently these requirements.

Finally, implementing *machine learning algorithms suited to smart city data and real-time processing* is another major challenge. Handling city-wide data introduces additional complexity, especially when considering relations between different data types and sensing devices. Current mobile devices have enough processing power to handle a broad set of use-cases, especially when dealing with data from integrated sensors (e.g., [19] uses on-device processing to classify urban noise sources). This could also be utilized as a means to enhance privacy, since processing would be performed locally, without requiring sensitive data to be uploaded to the cloud.

*Use cases*

We now proceed with a set of characteristic use cases, essentially highlighting our vision for our system and insights to the aforementioned challenges.

*IoT sensors to create better running and biking routes*: This use-case utilizes mobile and smartphone/smartwatch sensors to monitor environmental parameters to infer better routes for running and biking in terms of healthy environmental conditions. Parameters that could be sensed include air quality, noise pollution, pollen concentration, condition of roads, etc. Machine learning techniques could be used to identify anomalies in the sensed data, such as high pollutant or particle concentrations, or locations with high noise levels. Alerts regarding such events could be sent by the system to participating end-users to quantify or validate such data through annotations. Another use of data annotation contributions could relate to the sentiments of participants for their surroundings.

*The soundtrack of the city*: The concept is to create the aural and noise level maps of cities. This includes the use of smartphones' microphones to record noise or distinctive sounds of the urban landscape. Participants could use data annotations to pinpoint street musicians, sounds from birds or other animals and their place in the landscape, or sounds from public spaces like train or bus stations, or city halls, etc. Machine learning techniques could be used to generate general classifications that could subsequently made more specific by end-users providing additional data annotations. Users could also add descriptions and their sentiments towards places and sounds.

*Smart city event correlations*: A diverse smart city IoT infrastructure could "record" the same event from different aspects; a traffic jam could take place at a certain point in time (traffic data), while creating certain side effects, such as noise from car horns or engines (noise data), unusual levels of pollution (air quality data), etc. Since this kind of data is being fed to the system with similar spatiotemporal characteristics, such anomalies can be detected and correlated on a first level, and then be validated by end-users to define additional correlations.

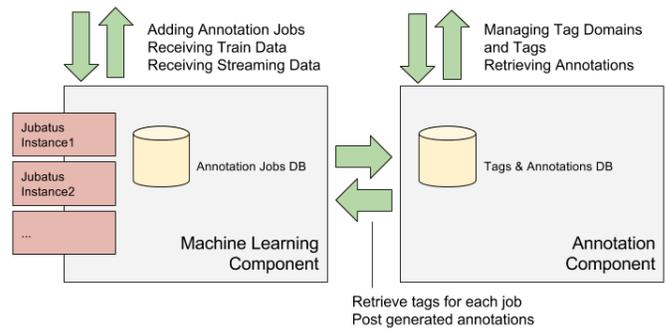

Figure 2: OrganiCity Annotation Service internal structure

ARCHITECTURE

OrganiCity federates existing smart city infrastructures, integrating urban data sources and services. Federated resources are exposed in this context, through a unified experimentation service and a central Context Broker. Our data annotation service is designed to operate on those unbounded incoming streams, to provide the additional knowledge required and increase their value and usefulness. *JAMAiCA* is capable of consuming, processing and annotating each individual data point to produce temporal annotations or nearby measurements to generate spatial annotations.

There are two main components (see Fig. 2) responsible for the whole annotation process:

The *Annotation Component* is responsible for maintaining a directory of all possible annotations in the form of *tags*. Tags are simple indicators of the annotated parameter, similar to the way tagging is performed in photos in social networks, or the use of hashtags in social status updates. *Tag domains* are created as collections of tags (e.g., "*high*", "*normal*" and "*low*") with a similar contextual meaning. Tag domains can be generic as those mentioned before or more application specific (e.g., the tag *"contains a beach"* for images). Users of the system can either select one of the tag domains already available, or create their own specifically for their application. Annotations are stored with additional information like numeric or text values. These entries can be user comments, a number that describes the abnormality of an observation, or a confidence indicator.

The *Machine Learning Component* orchestrates the machine learning process, including managing the executed jobs, training the instances with provided or retrieved data, and the exchange of real-time city data. Our system is capable of performing both anomaly detection and classification jobs over the streaming data. In both cases, after annotation jobs are added to the system the initial training data need to be submitted. After the initial training data are submitted, the annotation job starts with each data point examined and the result posted to the *Annotation Component*. The system is also agnostic of the actual machine learning process, as it capable of using multiple external services to identify extreme values or to classify data. This gives flexibility to experiment with machine learning algorithms and expandability to provide extra functionality in the future.

---

[1] https://www.w3.org/TR/annotation-protocol/

IMPLEMENTATION

In the rest of this section, we describe the technologies used for the implementation of our system and provide some initial results about the operation and the performance of the system.

In order to perform the analysis of the data, we use the Jubatus Distributed Online Machine Learning Framework [1]. *Jubatus* uses loose model sharing, a general computational framework for online and distributed machine learning. For each annotation process, we deploy a dedicated instance (either a *jubaanomaly* or *jubaclassifier*) and feed it with the provided training data. Our service communicates with each instance using RPC calls and with each call, the knowledge of the instance is enriched, as new data is feedback to the training mechanism in order to adapt its prediction and analysis engine. This setup allows us to horizontally scale on demand the machine learning infrastructure.

Data is fed to our system either directly or through an NGSI context broker [4]. More options like ActiveMQ or MQTT message queues can be implemented and then be added to the system. For our main use case, *JAMAiCA* uses a query context, provided during the creation of the annotation job, to register for updates on an NGSI context broker. This query acts as a set of selection parameters for the devices and sensors the job is interested in. In our implementation, we use the FIWARE Orion Context Broker as an input data source. After the subscription is established, Orion uses POST HTTP requests to notify our system of the newly received data following the NGSI specification. In the direct case, data need to be fed to the system via HTTP POST requests manually. The format of the HTTP body needs to be the formatted according to NGSI specification as well, for simplicity's sake.

For the development of both components, we use Java and the Spring Boot framework [5]. Spring Boot is Spring's convention-over-configuration solution for creating stand-alone, production-grade Spring based applications, as it simplifies the bootstrapping and development stages. It eases the process of exposing components, such as REST services, independently and offers useful tools for running in production, database initialization, environment specific configuration files and collecting metrics. In our case, we implemented both interfaces as RESTful web services. The API of the *Machine Learning Component* offers methods for handling primary HTTP requests (post, get, put and delete) that correspond to CRUD (create, read, update, and delete) operations on the *Annotation Jobs Database*, respectively. As a result, it allows experimenters to add and manage annotation jobs programmatically from their application. An annotation job can be either an anomaly detection or a classification process. Since the inception of the training process requires initial training data, additional methods that allow training a *Jubatus* instance for an existing annotation job are available. Lastly, a certain operation provides handling subscription updates from Orion or users (in the direct case) and starts the data validation against *Jubatus*. At the end of the training process, results are stored to the *Annotation Component*.

For the tag and annotation management in the *Annotation Component* we use Neo4j [6], a graph database that leverages data relationships and helps us build an intelligence around the tag domains and tags stored in our system. By traversing the relationships between the tags that comprise a tag domain, we can easily create suggestions for the appropriate tags a new annotation application may need to use. Also, the relationships between annotations be can used to extract higher knowledge for the cities or the experiments, especially when augmented with location and time information in order to identify events or situations that arise inside the cities. For example, an application can query for streets of the city where high atmospheric pollution and low vehicle speed is detected (indicating a possible traffic jam) and advise drivers to use alternative routes or means of transport when combined with information about the local subway timetables. The code of our system, as well as examples for its usage, is available on Github[2].

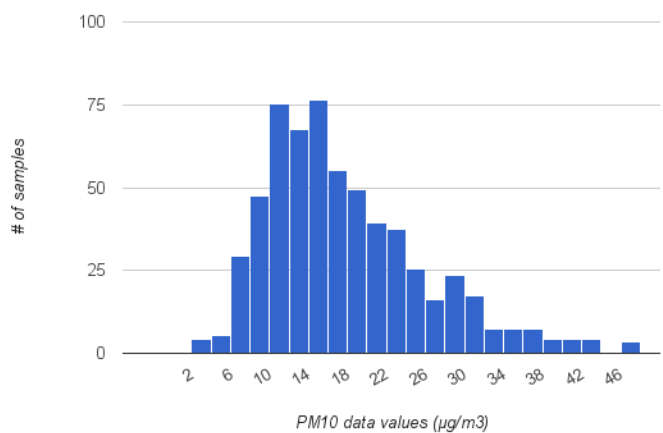

Figure 3: Histogram of the data used to train the Jubatus instance

*Results – Discussion*

To verify the performance of our system, we setup an anomaly detection job for analyzing atmospheric pollution in London, based on data for the particulate matter concentration ($PM_{10}$). As training data for our test case, we used data of the same area from the past 12 months (1000 nominal data points). We then let the system operate for 20 days, analyzing more than 40000 sensor measurements (translated to an average of 6 measurements every 15 minutes). The distributions of the values for the training data and the sensor data are presented in Fig. 3 and 4. A big part of the sensor measurements received is negative, pointing out a malfunction in the sensor devices, while another part of the measurements is greater than 50 μg/m$^3$, the level the European Union considers dangerous when exceeded for more than 35 days in a year. Our system was able to detect all values that were either negative or greater than the 50 μg/m$^3$ limit and record the time and date these values were abnormal.

---

[2] https://github.com/OrganicityEu/JAMAiCA

These 20 experimentation days helped us show that the data generated by smart city installations are not always trustworthy. A big number of the deployed sensor devices proved to be malfunctioning during our experiment (negative values), while a small number of measurements provided by the rest of the devices differed from the expected levels. The system performed without any problems for the whole duration of the experiment, on a virtual machine with limited resources (8GB of hard disk, 2GB RAM and 2 CPU cores) proving the system's scalability.

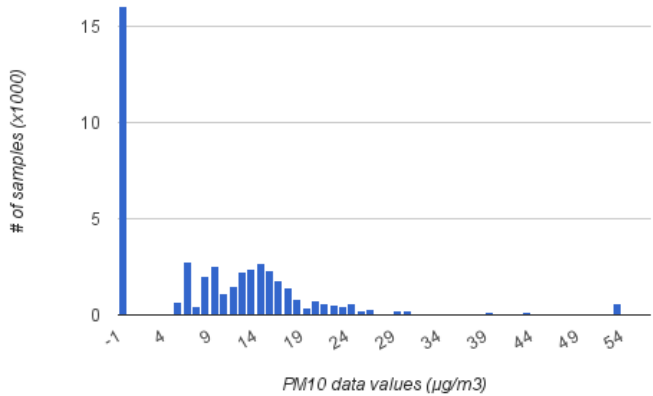

Figure 4: Histogram of sensor data received during the 20 experiment days.

## Conclusions – Future Work

In this article, we presented a solution that validates and enriches data produced by smart city infrastructures. We believe that this kind of processing is critical for IoT sensor data to become something more than simple datasets, i.e., a useful and reliable data source to facilitate the development of future city services. We presented a service capable of automatically detecting erroneous or unexpected sensor data using machine learning algorithms, classifying them and detecting real-world events and situations (in the form of annotation tags) based on provided training data sets. The initial tests we carried out showed that the proposed solution could effectively identify incidents with values from IoT devices in the London area diverging from the expected and nominal values (indicating either a malfunction or a dangerously high value).

To achieve the next level of smart city data analysis, we believe that citizen participation is of critical importance. Thus, in our next steps, we will focus on building interactive interfaces that make it easy for users to augment or confirm the automated annotations generated from our system, or provide their own input on non-annotated data. Finally, in order to further increase the participation rate and interest of the citizens, various methods for incentives and gamifications need to be assessed either in the context of rewards from the city to the citizens or benefits that city services could provide to the citizens directly.

## Acknowledgment

This work has been partially supported by the EU research project OrganiCity, under contract H2020-645198.

## References

[1] OrganiCity, Co-creating smart cities of the future, http://organicity.eu (retrieved online July 31st, 2016)

[2] Shohei Hido, Seiya Tokui, Satoshi Oda, "Jubatus: An Open Source Platform for Distributed Online Machine Learning" in *Neural Information Processing Systems*, 2013.

[3] Puiu, D., Barnaghi, P., Tönjes, R., Kümper, D., Ali, M. I., Mileo, A., & Gao, F., "CityPulse: Large Scale Data Analytics Framework for Smart Cities" in *IEEE Access*, 2016. doi:10.1109/ACCESS.2016.2541999

[4] NGSI Specification, (retrieved online July 31st, 2016) https://forge.fiware.org/plugins/mediawiki/wiki/fiware/index.php/FI-WARE_NGSI_Open_RESTful_API_Specification

[5] Spring Boot framework, http://projects.spring.io/spring-boot (retrieved online July 31st, 2016)

[6] Jim Webber, "A programmatic introduction to Neo4j" in *Proceedings of the 3rd annual conference on Systems, Programming, and Applications: Software for Humanity* (SPLASH '12), 2012. doi: 10.1145/2384716.2384777

[7] CKAN, http://ckan.org/ (retrieved online July 31st, 2016)

[8] Socrata, https://socrata.com/ (retrieved online July 31st, 2016)

[9] F. Ganz, P. Barnaghi & F. Carrez, "Automated Semantic Knowledge Acquisition From Sensor Data," in *IEEE Systems Journal*, 2014. doi: 10.1109/JSYST.2014.2345843

[10] Guo, B., Wang, Z., Yu, Z., Wang, Y., Yen, N., Huang, R., & Zhou, X., "Mobile crowd sensing and computing: The review of an emerging human-powered sensing paradigm" in *ACM Computing Surveys*, 2015 doi: 10.1145/2794400

[11] Luis Sánchez et al., "SmartSantander: IoT experimentation over a smart city testbed" in *Computer Networks*, 2014.

[12] Barowy, D. W., Curtsinger, C., Berger, E. D., & McGregor, A., "Automan: A platform for integrating human-based and digital computation" in *ACM SIGPLAN Notices*, 2012. doi: 10.1145/2927928

[13] Chilton, L. B., Little, G., Edge, D., Weld, D. S., & Landay, J. A., "Cascade: Crowdsourcing taxonomy creation" in *Proceedings of the SIGCHI Conference on Human Factors in Computing Systems*, 2013. doi: 10.1145/2470654.2466265

[14] Knerr, Torben. (2006). Tagging ontology-towards a common ontology for folksonomies (retrieved online July 31st, 2016) http://tagont.googlecode.com/files/TagOntPaper.pdf

[15] CitySDK-City Service Development Kit, http://www.citysdk.eu/, (retrieved online July 31st, 2016)

[16] Bischof, S., Karapantelakis, A., Nechifor, C., Sheth, A., Mileo, A., & Barnaghi, P. Semantic modelling of smart city data.

[17] Raykar, V. C., Yu, S., Zhao, L. H., Valadez, G. H., Florin, C., Bogoni, L., & Moy, L., "Learning from crowds" in *Journal of Machine Learning Research*, 2010.

[18] Welinder, P., Branson, S., Perona, P., & Belongie, S. J., "The multidimensional wisdom of crowds" in *Advances in neural information processing systems*, 2010.

[19] Sounds of New York City project, https://wp.nyu.edu/sonyc/ (retrieved online July 31st, 2016)